\PassOptionsToPackage{dvipsnames}{xcolor}
\documentclass[10pt,conference]{IEEEtran}
\IEEEoverridecommandlockouts

\usepackage[utf8]{inputenc}
\usepackage[T1]{fontenc}
\usepackage{svg}
\usepackage[ruled,vlined]{algorithm2e}
\usepackage[normalem]{ulem}
\usepackage{amsmath,amssymb,amsfonts}
\usepackage{graphicx}
\usepackage{booktabs}
\usepackage{multirow}
\usepackage{adjustbox}
\usepackage{url}
\usepackage{hyperref}
\usepackage{subcaption}  
\usepackage{soul}
\usepackage{threeparttable}
\usepackage{siunitx}
\usepackage{algorithmic}
\usepackage{listings}
\usepackage[most]{tcolorbox}
\usepackage{tikz}
\newcommand{\spice}{\raisebox{-0.1em}{\includegraphics[height=1em]{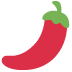}}}

\usepackage{xspace}
\usepackage{soul}
\setuldepth{Berlin}
\usepackage{hyperref}
\hypersetup{colorlinks=true,breaklinks=true}
\usepackage{multirow}
\usepackage{makecell}
\usepackage{balance}
\usepackage{enumitem}
\usepackage{pifont}
\usepackage{float}
\usepackage{csquotes}
\usepackage[numbers,sort&compress]{natbib}
\usepackage{fixltx2e}

\newcommand{\tinyskip}{\vspace{1pt}}

\newboolean{showcomments}
\setboolean{showcomments}{true}

\ifthenelse{\boolean{showcomments}}
{\newcommand{\nbc}[3]{
 {\colorbox{#3}{\bfseries\sffamily\scriptsize\textcolor{white}{#1}}}
 {\textcolor{#3}{\sf\small$\blacktriangleright$\textit{#2}$\blacktriangleleft$}}
 }
}
{\newcommand{\nbc}[3]{}
 }

\newcommand\gopi[1]{\nbc{Gopi}{#1}{red}}
\newcommand\aadi[1]{\nbc{Aadi}{#1}{blue}}

\newcommand\todo[1]{\nbc{TODO}{#1}{orange}}

\definecolor{box-blue}{RGB}{25,60,130}
\definecolor{box-gray}{RGB}{248,248,248} 

\newtcbtheorem
  [no counter]
    {Summary}{\textbf{\small{Summary}}}
  {
    enhanced,
    breakable,
    colback=box-gray,
    colframe=box-blue,
    coltitle=white,
    colbacktitle=box-blue,
    boxrule=0.8pt,
    arc=4pt, outer arc=4pt,
    top=6pt, bottom=2pt, left=6pt, right=6pt,
    before skip=10pt,
    after skip=10pt,
    drop shadow={shadow xshift=1pt,shadow yshift=-1pt,opacity=0.15},
    attach boxed title to top left={yshift=-2mm, xshift=3mm},
    boxed title style={enhanced,boxrule=0pt,arc=2pt},
    colupper=box-blue,                           
  }{summary}

\newcommand{\rqone}{RQ1: How accurately does SPICE reproduce expert-curated SWE-L labels?}
\newcommand{\rqtwo}{RQ2: How similar are the SPICE-generated rationales to those in SWE-L?}
\newcommand{\rqthree}{RQ3: What are the cost considerations associated with SPICE?}

\begin{document}

\title{\textbf{SPICE}\spice: An Automated \textbf{S}WE-Bench Labeling \textbf{P}ipeline for \textbf{I}ssue Clarity, Test \textbf{C}overage, and \textbf{E}ffort Estimation}

\author{%
  \IEEEauthorblockN{%
    Gustavo A. Oliva\textsuperscript{*}\IEEEauthorrefmark{3},\,
    Gopi Krishnan Rajbahadur\textsuperscript{*}\IEEEauthorrefmark{3},\,
    Aaditya Bhatia\textsuperscript{*}\IEEEauthorrefmark{3},\,
    Haoxiang Zhang\IEEEauthorrefmark{2}\\
    Yihao Chen\IEEEauthorrefmark{2},\,
    Zhilong Chen\IEEEauthorrefmark{2},\,
    Arthur Leung\IEEEauthorrefmark{2},\,
    Dayi Lin\IEEEauthorrefmark{2},\,
    Boyuan Chen\IEEEauthorrefmark{2},\,
    Ahmed E. Hassan\IEEEauthorrefmark{3}%
  }%
  \vspace{1ex}
  \IEEEauthorblockA{\IEEEauthorrefmark{2}Center for Software Excellence, Huawei Canada\\
  }
  \IEEEauthorblockA{\IEEEauthorrefmark{3}Software Analysis and Intelligence Lab (SAIL), School of Computing, Queen’s University, Canada\\
  Emails: \{aaditya.bhatia,gustavo,ahmed\}@cs.queensu.ca, 
  grajbahadur@acm.org
  }
  
  \{haoxiang.zhang,yihao.chen,zhilong.chen,arthur.leung1,dayi.lin,boyuan.chen1\}@huawei.com
  \thanks{\textsuperscript{*}First three authors contributed equally.}
}
\maketitle

\begin{abstract}
  High-quality labeled datasets are crucial for training and evaluating foundation models in software engineering, but creating them is often prohibitively expensive and labor-intensive. We introduce SPICE, a scalable, automated pipeline for labeling SWE-bench-style datasets with annotations for issue clarity, test coverage, and effort estimation. SPICE combines context-aware code navigation, rationale-driven prompting, and multi-pass consensus to produce labels that closely approximate expert annotations. SPICE's design was informed by our own experience and frustration in labeling more than 800 instances from SWE-Gym. SPICE achieves strong agreement with human-labeled SWE-bench Verified data while reducing the cost of labeling 1,000 instances from around \$100,000 (manual annotation) to only \$5.10. These results demonstrate SPICE's potential to enable cost-effective, large-scale dataset creation for SE-focused FMs. To support the community, we release both SPICE\spice tool and SPICE\spice Bench, a new dataset of 6,802 SPICE-labeled instances curated from 291 open-source projects in SWE-Gym (over 13x larger than SWE-bench Verified). 

\end{abstract}


\section{Introduction}\label{sec:intro}

In the era of Foundation Models (FMs) like Large Language Models (LLMs), high-quality software engineering (SE) datasets fulfill two key roles: (i) they benchmark model performance on real-world SE tasks and (ii) serve as valuable corpora for pretraining and fine‑tuning FMs. 

A prominent example of a high-quality SE dataset is SWE-bench (\textbf{S}oft\textbf{W}are \textbf{E}ngineering \textbf{bench}mark) Verified~\cite{swebenchverified2024} (\textbf{SWE-V}).
SWE-V was designed to assess an FM's ability to produce a valid code patch that fixes an issue report from a given software repository. More specifically, each instance (row) in SWE-V represents a code repair task and contains three key fields: the issue description (title and body), a reference (golden) solution patch, and corresponding test cases used for evaluation. We refer to similar SE datasets containing these key fields as \textit{SWE-bench-like datasets}. Notable SWE-bench-like datasets include SWE-Gym~\cite{pan2024training}, Multi-SWE‑bench~\cite{zan2025multi}, and SWE-bench Multimodal~\cite{yang2024swe}. The tasks in SWE-V span 12 real-world Python projects, including widely used ones such as Django.

In terms of role (i), SWE-V has been extensively used to benchmark FMs as well as agentic scaffolds built on top of those. A leaderboard is maintained and constantly updated~\citep{swebench2025leaderboard}. When major players release a new FM (or FM version), they typically advertise their score on SWE-V or one of its variations. That is, SWE-V has become the de facto SE benchmark. In terms of role (ii), researchers have leveraged the rich code and problem solving logic in SWE-V to train and improve models as well as create autonomous agents. A prominent example is SkyRL~\cite{cao2025skyrl}, which is an open-source reinforcement learning framework developed by the Berkeley Sky Computing Lab to train AI agents for long-horizon tasks in real-world environments. SkyRL leverages SWE‑V as both the environmental scaffold and training signal for reinforcement learning. Beyond SkyRL, recent evidence shows that SWE-bench-like datasets can even enhance the reasoning capabilities of general-purpose FMs for everyday tasks~\cite{dickson2024code,zhang2025unveiling}. 

While SWE-bench-like datasets have vast practical value, they are also exceptionally costly and difficult to construct. SWE-V was created by OpenAI as a response to the original SWE-bench, which had incomplete issue reports, issue reports containing the reference solution, and unit tests checking for aspects not even described in the issue. These data problems are critical as they lead to \textit{unsolvable issues}, since either the problem to be solved cannot be properly understood (not even by a human) or candidate solutions generated by the model cannot be adequately verified (e.g., a valid solution may fail the tests). SWE-V excludes unsolvable tasks from the original dataset, focusing solely on those that are verifiably solvable. This was achieved through manual and careful labeling of issues (\textit{are issues clear enough?}) and test patches (\textit{is test coverage adequate and precise?}) of 1,699 randomly drawn instances from SWE-bench. We refer to this dataset as \textbf{SWE-L}. We estimate that constructing SWE-V (500 instances) required approximately 2,265.3 engineer hours of manual labeling. The labeling work was performed by professional software developers hired specifically for the task.

Beyond the significant (and often prohibitive) cost of expert labelers, further challenges include the difficulty in training them for each code repository (e.g., each repository tackles a different application domain) and the risk of potential labeler bias~\cite{carletta1996assessing,cohen1960coefficient}. While a scientific labeling process involves a reasonable Inter-Rater Agreement (IRA) for the labeling results to be trustworthy, obtaining a high IRA when labeling complex software artifacts is hard in practice. Indeed, by inspecting the public annotation data made available by OpenAI, we observed a Krippendorff's~$\alpha$~\cite{krippendorff2011computing} IRA of just 0.24 for issue clarity labeling and 0.41. for test adequacy labeling. Our own manual efforts of labeling a SWE-bench-like dataset the rubric from SWE-V revealed similarly low IRAs in the range of 0.0-0.4, much below the minimum acceptable value of 0.64 (see Section~\ref{sec:pre-study}).

In summary, manual labeling of SWE-bench-like datasets is time-consuming, expensive, subject to labeler bias, and does not scale. Compounding these challenges, the scarcity of high-quality, manually labeled benchmarks has driven the community to rely repeatedly on SWE-V, considerably raising the risks of overfitting and benchmark data leakage ~\cite{zhou2023don} (i.e., models being evaluated with data that they have seen during pretraining). Consequently, it is crucial to establish a streamlined approach for the rapid and reliable generation of high-quality SWE-bench-like datasets. To address these challenges, we present \textbf{SPICE}\spice: an automated  platform for labeling SWE-bench-like datasets that strives for data quality at scale. Leveraging Aider~\cite{Aider2025}, a context-aware code navigation and editing AI assistant, SPICE extracts relevant information from repositories, constructs structured prompts, and synthesizes high-quality labels for issue clarity and test coverage in a scalable manner. To enhance robustness, SPICE labels each instance multiple times and subsequently performs an aggregation through majority voting (a.k.a., self-consistency via stochastic sampling~\citep{wang2023selfconsistencyimproveschainthought}). This process yields a more statistically grounded approximation of expert consensus. 

The development of SPICE was informed by our own experience in manually labeling a SWE-bench-like dataset (SWE‑Gym~\cite{pan2024training}). Over the course of annotating over 800 instances, we encountered persistent subjectivity, low IRA, and significant cognitive and time demands, even with a team of expert annotators (see Section~\ref{sec:pre-study}). These challenges made clear that manual labeling at scale is neither sustainable nor reproducible. SPICE emerges from this hard-earned experience as a principled response. As a demonstration of SPICE's scalability, we release SPICE\spice-bench---a curated dataset of 6,802 automatically labeled instances drawn from 291 real-world open-source projects in SWE-Gym~\cite{pan2024training}. This is the largest known collection of \textit{solvable} SWE-bench-like tasks, and is over 13 times larger than SWE-V's human-labeled subset. SPICE-bench provides a rich resource for fine-tuning, benchmarking, and training SE-focused foundation models.

In this paper, we evaluate SPICE's effectiveness by demonstrating its scalability and consistency along the following research questions (RQs):
\begin{itemize}[leftmargin=*]
    \item \textbf{\rqone} \\ 
    \textit{SPICE achieves an accuracy of 67\% in identifying the instances that were labeled as ``unsolvable'' in SWE-L.}
    \item \textbf{\rqtwo} \\ 
    \textit{The rationales produced by SPICE's labelers are semantically similar to those of expert labelers and are sufficiently detailed to be debuggable.}
    \item \textbf{\rqthree} \\ 
    \textit{In its default setting, SPICE costs only \$5.10 to label 1,000 instances\footnote{All reported monetary values in this paper are in US dollars (USD).}.}
\end{itemize}

\tinyskip The contributions of this paper can be summarized as follows:
\begin{itemize}[wide = 0pt]
    \item We release a labeling tool that can be adopted and extended by the open source community\textsuperscript{*}.
    \item We release a dataset of 6,802 SPICE-labeled, which is the \emph{largest} curated collection of \textit{solvable} SWE-bench-like instances till date. Our dataset is available on Hugging Face as SPICE-bench\textsuperscript{*}\footnote{\textsuperscript{*}The URL will be made available after acceptance, but the dataset and our SPICE\spice tool is enclosed with the submission as supplementary material}.
    \item We present an experience report on labeling complex software artifacts, highlighting key pain points in the manual labeling process. These insights can be highly valuable to both academic and industry practitioners involved in manual labeling.
\end{itemize}

\section{SWE-Bench Verified (SWE-V)}
\label{sec:background}

Introduced in late 2023, SWE-bench~\cite{jimenez2023swe} marked a critical shift in how we evaluate FMs, moving from standalone code generation tasks (e.g., HumanEval~\cite{humaneval} and MBPP~\cite{mbpp}) to realistic, repository-level code maintenance scenarios. Each instance in SWE-bench pairs a real issue (typically a GitHub issue) with the corresponding pull request that resolved it, grounded in the full project repository. Models must modify the codebase to fix the issue such that newly added tests will pass (\texttt{FAIL\_TO\_PASS}) while all existing tests continue to succeed (\texttt{PASS\_TO\_PASS}). This formulation mirrors real-world workflows: correctness is verified through test execution, changes span multiple files, and solutions depend on long-context reasoning across a live codebase. 

SWE-V was introduced in mid-2024 as a solution to SWE-bench's validity issues~\cite{swebenchverified2024}. This subset of 500 instances was selected through an intensive labeling campaign aimed at removing the flawed instances (unsolvable issues) ~\cite{aleithan2024swebenchenhancedcodingbenchmark}. The campaign recruited a total of 93 professional Python developers, who labeled 1,699 random instances from SWE-bench. We refer to this dataset as \textbf{SWE-L} (SWE-Bench labeled). Every instance was labeled by 3 developers. More specifically, labelers were assigned two fundamental assessment tasks: \textbf{Issue Clarity Assessment (ICA)} and \textbf{Test Coverage Assessment (TCA)}. An instance was deemed \textit{verified} when it passed \textit{both} assessments. Additionally, as supplementary information (i.e., not used for dataset filtering), labelers were asked to estimate how long it would take an experienced software engineer, after spending a few hours getting familiar with the codebase, to write a patch that resolves the issue. Since this effort estimation does not impact the filtering, we do not discuss it in this paper (however, SPICE does provide an estimation for effort as well). In the following, we describe ICA and TCA. More details can found in SWE-V webpage~\cite{swebenchverified2024}.

\tinyskip\noindent\textbf{Issue Clarity Assessment (ICA).} The FM under evaluation is expected to produce a patch based on the provided the issue description (title and body) and codebase. If the issue description is vague or poorly defined, generating a patch that effectively addresses the issue can become significantly more difficult or even impossible in some cases. Given solely an issue description (i.e., no access to external links nor the issue discussion thread), labelers were requested to assess issue clarity using the following ordinal scale:
\begin{tcolorbox}[colback=gray!10, colframe=gray!50, boxrule=0.3pt, arc=2pt, left=4pt, right=4pt, top=2pt, bottom=2pt]
\begin{small}
\begin{itemize}[wide = 0pt, topsep = 3pt, itemsep = 1pt]
    \item \textbf{0 (Well-specified):} Clear requirements; minimal ambiguity.
    \item \textbf{1 (Minor ambiguity):} Some details are missing but an experienced developer can infer a sensible solution.
    \item \textbf{2 (Significant ambiguity):} Vague description, multiple interpretations possible.
    \item \textbf{3 (Poorly specified): }Nearly impossible to derive meaningful solutions without additional clarification.
\end{itemize}
\end{small}
\end{tcolorbox}
\tinyskip\noindent\textbf{Test Coverage Assessment (TCA).} To assess an FM's proposed solution, its generated patch is applied to the codebase, followed by the execution of the FAIL\_TO\_PASS and PASS\_TO\_PASS sets of unit tests. If the patch is applied without errors and all tests pass, the solution is deemed to have successfully addressed the problem. Given the issue description, the gold patch, and the test patch (and the possibility of navigating through the codebase), labelers were requested to assess test coverage using the following ordinal scale:
\begin{tcolorbox}[colback=gray!10, colframe=gray!50, boxrule=0.3pt, arc=2pt, left=4pt, right=4pt, top=2pt, bottom=2pt]
\begin{small}
\begin{itemize}[wide = 0pt, topsep = 3pt, itemsep = 1pt]
    \item \textbf{0 (Perfect coverage):} Tests cover all reasonable solutions.
    \item \textbf{1 (Good coverage):} Tests cover most solutions, though some atypical solutions may be missed.
    \item \textbf{2 (Limited coverage):} Tests overlook several reasonable solutions.
    \item \textbf{3 (Poorly scoped):} Tests are excessively narrow or overly broad, significantly misaligned with the issue requirements.
\end{itemize}
\end{small}
\end{tcolorbox}
\tinyskip\noindent\textbf{Filtering-in instances (verification criteria).} For benchmark instances to be filtered-in, two conditions must be simultaneously met: (i) the issues are well-specified enough (ICA < 2) and (ii) the test coverage is adequate enough (TCA < 2).  
\section{From Manual Effort to Automation: Lessons from Replicating SWE-bench-Verified}
\label{sec:pre-study}


SWE-V has emerged as the de facto standard for evaluating FMs on realistic software maintenance tasks. However, it remains unclear whether its labor-intensive annotation protocol is replicable by independent teams. To tackle this problem, we conducted a manual annotation campaign modeled closely after SWE-V's rubric, targeting 840 unlabeled instances from the SWE-Gym dataset - an unlabeled, SWE-bench-like dataset - to curate our own dataset.

In addition to assessing replicability, we also focused on extracting actionable lessons, evaluating the costs and limitations of manual curation, and determining whether scaling such efforts was feasible (or whether we needed to invest in building an automated labeling solution). Given that both paths required significant resource investment, it was critical to evaluate whether verification of SWE-bench-like datasets could be conducted both rigorously and efficiently.
\subsection{Goals and Context}
Our goals were threefold:

\begin{itemize}[wide = 0pt]
\item \textbf{Assess Generalizability:} Determine whether the data quality issues addressed by SWE-V were specific to its original 12 Python projects or indicative of broader structural challenges in SWE-bench-like datasets.

\item \textbf{Validate Annotation Protocols:} Reapply OpenAI's labeling rubric focused on Issue Clarity Assessment (ICA) and Test Coverage Assessment (TCA), to test whether the criteria are practically operationalizable by third-party experts. A useful benchmark must be not only accurate, but also replicable.

\item \textbf{Establish Baselines for Automation:} If manual curation proved infeasible under our resource constraints, we aimed to use this effort to generate empirical benchmarks on cost, subjectivity, and inter-rater agreement (IRA) to guide the development and evaluation of automated tools like SPICE.
\end{itemize}

Our labeling team consisted of seven software engineering researchers: the first four authors of this paper (PhD-level or senior graduate students in SE) and three external annotators who hold a PhD in SE. Each member of the team also had at least three years of professional software development experience. Annotations were conducted within an applied research environment at an industrial lab, where time, personnel, and infrastructure were constrained by real-world limitations. These included narrow annotation windows, tight deliverable timelines, and a prioritization of practical tool development over long-form qualitative studies. While this context limited our ability to fully adhere to methodological standards (e.g., grounded theory~\cite{urquhart2013grounded}), it closely reflects the conditions under which organizations might realistically pursue SWE-bench-like dataset curation at scale.

\subsection{Method}
\tinyskip \noindent\textbf{Inter-Rater Agreement (IRA) Calculation.}
When computing IRA between our annotators for both ICA and TCA, to assess whether the labels can be trusted and reliably used, we do not evaluate agreement directly on the raw 4-point annotations. Instead, we discretize them as follows: 

\setlength{\abovedisplayskip}{0pt}
\begin{small}
\begin{equation}
\text{ICA\textsubscript{label}}(score) =
\begin{cases}
\texttt{Well-specified}, & \text{if } score \in \{0, 1\} \\
\texttt{Underspecified}, & \text{if } score \in \{2, 3\}
\end{cases}
\end{equation}
\begin{equation}
\text{TCA\textsubscript{label}}(score) =
\begin{cases}
\texttt{Adequate}, & \text{if } score \in \{0, 1\} \\
\texttt{Inadequate}, & \text{if } score \in \{2, 3\}
\end{cases}
\end{equation}
\end{small}

We adopt this mapping to mirror SWE-V's own downstream use of these labels~\cite{swebenchverified2024}. Since our goal is to assess whether our pipeline can approximate SWE-V quality, we align with their discretization scheme.

For all IRA comparisons reported in this section and throughout the study, we use Krippendorff's~$\alpha$~\cite{krippendorff2011computing}, which satisfies three essential criteria: (1) it supports any number of annotators, (2) it handles nominal data, and (3) it is widely trusted in empirical research~\cite{bhatia2023towards,fabbri2021summeval}. Unlike Cohen's Kappa or Fleiss' Kappa, which may produce misleadingly low values in the presence of class imbalance, Krippendorff's~$\alpha$ accommodates such imbalance and supports a range of data types including nominal, ordinal, interval, and ratio.

We interpret Krippendorff's~$\alpha$ values using the thresholds recommended by~\citet{krippendorff2018content}:

\setlength{\abovedisplayskip}{0pt}
\begin{small}
\begin{align*}
\text{Krippendorff's }\alpha =
    \begin{cases}
        \text{Poor,} & \alpha < 0.66 \\
        \text{Moderate,} & 0.66 \le \alpha < 0.80 \\
        \text{Substantial,} & 0.80 \le \alpha \\
    \end{cases}
\end{align*}
\end{small}

\tinyskip\noindent\textbf{Initial Calibration and Pilot Phase.}  
We began with a one-week calibration phase, systematically reviewing publicly available SWE-V annotations and rationales~\cite{swebenchverified2024} to develop a shared understanding of the annotation criteria. During daily meetings, we resolved conflicts, refined our labeling guidelines, and clarified edge cases and ambiguities. Annotators also received extensive support through FM-generated acclimatization documents\footnote{An example of our acclimatization can be found \href{https://anonymous.4open.science/r/ASE_Submission_SPICE-D083/}{here}} summarizing each repository's structure, architecture, and test practices.


    \begin{figure}[!ht]
      \centering\includegraphics[width=0.8\columnwidth,keepaspectratio]
      {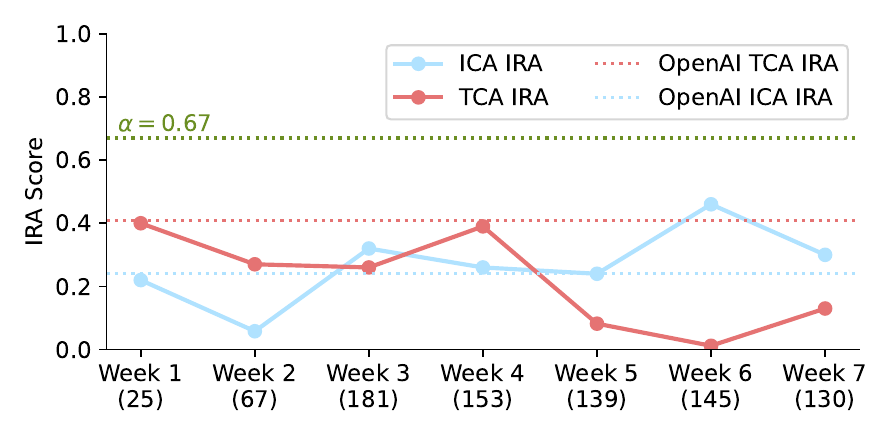}
      \caption{Krippendorff's $\alpha$ for the main labeling phase. The green line of $\alpha$ is the minimum acceptable level for the labeling results to be meaningful. The numbers below each week show how many instances we labeled each week.}
      \label{fig:prestudy_IRA}
    \end{figure}
\tinyskip\noindent\textbf{Main Annotation Phase}  
Following the initial pilot phase, we annotated 840 instances over seven weeks. Each instance was assigned to two annotators, who labeled both ICA\textsubscript{score} and TCA\textsubscript{score}. Each instance typically took anywhere between 20 to 40 minutes to annotate. When switching to a new project, annotators often required up to two hours to get acclimatized. We calculated IRA for both ICA and TCA tasks weekly and held at least one meeting per week to discuss conflicts and annotation challenges. Figure~\ref{fig:prestudy_IRA} presents the weekly IRA scores for both tasks.

\subsection{Results, Lessons Learned, and Key Takeaways}

\tinyskip\noindent\textbf{Lesson 1: Too much pain for too little gain---Creating a SWE-V-like dataset through manual curation is extremely difficult with small annotation teams.}  
Over seven weeks, we labeled 840 instances, investing approximately \textbf{882 person-hours} (approx. 22 engineer-weeks). Despite this effort, Krippendorff's~$\alpha$ never exceeded the reliability threshold of $\alpha = 0.67$ (green line in Figure~\ref{fig:prestudy_IRA}). While OpenAI used a moderator to resolve disagreements, incorporating such a process would introduce an additional 25\% overhead, making it prohibitively expensive. Although we did not record metrics, all labelers reported significant burnout due to the high cognitive demand and unrewarding nature of the task. Ultimately, the manual effort simply did not justify the return.

\tinyskip\noindent\textbf{Key Takeaway.} Without a large annotation team or a much longer timeline, manually curating a SWE-V-like dataset at scale is infeasible.

\tinyskip\noindent\textbf{Lesson 2: Beauty is in the eye of the beholder---and so are ICA and TCA scores.}  
IRA varied significantly across projects and weeks. ICA\textsubscript{IRA} ranged from \textbf{0.24} to \textbf{0.46}, while TCA\textsubscript{IRA} spanned from negative (\textbf{-0.012}) to a modest \textbf{0.39}. These values mirror the low IRAs of labelers in SWE-V (ICA\textsubscript{SWE-V-IRA} = 0.21 and TCA\textsubscript{SWE-V-IRA} = 0.41), highlighting the universal difficulty of the task. Persistent disagreements on what constitutes an \texttt{adequate} test or a \texttt{well-specified} issue made consistency difficult to achieve. These observations align with findings from~\citet{borstler2023developers}, who noted that developers often hold divergent views on what constitutes quality code, whereas, in our case, what qualifies as adequate tests or well-specified issues. 

\tinyskip\noindent\textbf{Key Takeaway.} Defining what makes an issue \texttt{well-specified} or a test \texttt{adequate} is inherently subjective and difficult to standardize.

\tinyskip\noindent\textbf{Lesson 3: Each project is a universe unto itself – and navigating multiple universes takes (too much) time.}  
Figure~\ref{fig:prestudy_IRA} shows a deceptively higher TCA\textsubscript{IRA} for Week 1. This occurred because many SWE-Gym instances we annotated that week came from the \textit{pydantic} project, a project we had previously discussed in depth while reviewing \textit{non-overlapping} instances from SWE-V. This familiarity gave annotators a head start. However, as the labeling expanded to other repositories, agreement deteriorated. The \textit{SQLGlot} project in Week 4 led to the lowest TCA\textsubscript{IRA}, as annotators lacked the deep domain knowledge needed to evaluate test coverage effectively. Even with two-hour onboarding sessions and FM-generated acclimatization documents, repository-specific context demands far more effort than such resources can provide.

\tinyskip\noindent\textbf{Key Takeaway.} Without involvement from developers who actively contribute to each project, large-scale manual labeling for SWE-bench-like datasets is impractical.

All of these lessons underscored the need for an automated solution that could deliver rubric-aligned, interpretable labels at scale and minimal cost. Importantly, we realized that extremely high accuracy was not the primary requirement. Instead, consistency, scalability, cost-efficiency, and the ability to audit and debug model-generated labels were far more critical. These realizations shaped the design goals of SPICE, which aims to serve as a practical, extensible alternative to manual labeling of SWE-bench-like benchmarks.

\section{SPICE~\spice}
\label{sec:tool}

SPICE\spice~primarily consists of two modular pipelines: the \textit{Issue Clarity Assessment (ICA) pipeline} and the \textit{Test Coverage Assessment (TCA) pipeline}. Together, they automate the SWE-V protocol for evaluating issue clarity and test adequacy. Each pipeline runs three times and SPICE applies majority voting to determine the final label (median is used in cases where a majority does not exist). Below, we describe the design, inputs, and outputs of each pipeline.

\subsection{Issue Clarity Assessment (ICA) Pipeline}
The ICA pipeline takes the \texttt{issue title} and \texttt{issue description} of a GitHub issue and classifies that issue as either \textit{well-specified} or \textit{underspecified}, using a \textit{rationale-informed prompt} that generates both a label and a supporting explanation. We first preprocess each issue by stripping markdown and retaining only relevant content, following~\citet{bai2024special} to avoid misleading the FM with formatting artifacts. We then pass the cleaned issue to a FM using our rationale-informed prompt.

\tinyskip
\noindent\textbf{Rationale-Informed Prompt.}  
Standard prompting strategies such as zero-shot, few-shot, chain-of-thought, and domain informed checklist-based approaches failed to consistently produce high-quality issue clarity labels. Inspired by~\citet{lin2024prompt}, who showed that human feedback improves prompt effectiveness, we grounded our prompt in developer rationales provided by SWE-L annotators.

We selected OpenAI-labeled issues where all three annotators agreed on the label. From this filtered pool, we performed \textbf{stratified sampling} ($n=5$ per clarity level; see Section~\ref{sec:background}) to capture a diverse set of 20 rationales across issues with different clarity levels. For each selected issue, we extracted the \texttt{title}, \texttt{description}, \texttt{label}, and all three \texttt{rationales} provided by the SWE-L labelers (see Section~\ref{sec:background}). We analyzed these with DeepSeek-R1~\cite{deepseekr1} using exploratory prompting to identify recurring rationale patterns, such as missing context, vague language, or lack of reproducibility and how they related to an issue being labeled \textit{well-specified} or \textit{underspecified}. These patterns helped us create a \textbf{rationale checklist} that guided prompt construction.

We then constructed our rationale-informed prompt by iteratively refining this checklist using a prompt that encouraged the FM to emulate these reasoning behaviors without relying on explicit few-shot examples. We found that including examples often confused the FM, as finding generalizable examples was difficult. We evaluated successive versions of the prompt on the same 20 issues used to construct the checklist, measuring both label accuracy and rationale quality. We selected the final prompt based on these evaluations. These 20 issues are excluded from all other parts of our study. 

\tinyskip\noindent\textbf{Candidate Solution Identification}  
We also extended the ICA prompt to detect whether the issue report includes a \textit{candidate solution}, such as code snippets, outlines, or textual hints that suggest how to fix the problem.

\subsection{Test Coverage Assessment (TCA) Pipeline}

Labeling tests require a deep understanding of the codebase. To achieve this goal and avoid ``reinventing the wheel'', we use Aider~\cite{Aider2025}. Aider is an open-source AI pair programming tool that integrates directly into the developer's terminal (shell). A key component of Aider is its \textit{RepoMap} algorithm~\citep{AiderRepomap}. This algorithm creates a concise representation of an entire Git repository (via \texttt{ctags} and \texttt{Tree-sitter}), capturing key classes and functions along with their types and call signatures. This enables precise understanding of the code being edited and its connections within the broader codebase. For large repositories, a graph ranking algorithm is used to identify and include only the most relevant parts of the codebase within a FM's token constraints, optimizing efficiency and effectiveness. Aider will adjust this token quota based on the chosen FM.

The TCA pipeline works in four simple steps. First, a structured prompt is constructed to describe the TCA task (see Section~\ref{sec:background}). Second, to help Aider narrow down the relevant context, we instruct it to focus on the files included in the \texttt{gold patch} and the \texttt{test patch} by using the \texttt{/read-only (file)} command. Third, we execute the structured prompt using Aider's \texttt{/run (prompt)} command. Internally, Aider appends a concise representation of the repository as context to the prompt and calls the underlying FM. Finally, we take Aider's output and prompt an auxiliary model (GPT-4o mini by default) in order to parse that output and retrieve the test label score.

We note that Aider is a terminal tool and not an API. Therefore, we implemented a thin python API to interact with Aider in a programmatic way (Aider is made available as a python package called \texttt{Aider-chat}). SPICE exposes key Aider parameters, such as the FM choice and the token quota for RepoMap.

\section{SPICE Evaluation Results}\label{sec.results}

\subsection{\rqone}

\noindent\textbf{Motivation.} For SPICE to be a viable alternative, labels predicted by SPICE must be at least as reliable as those produced by human experts. Therefore, in this RQ, we evaluate how accurately SPICE can replicate expert-level judgments for ICA and TCA.


\noindent\textbf{Approach.} To evaluate how accurately SPICE reproduces expert-curated SWE-L labels, we adopt a two-pronged strategy. First, we perform an automatic evaluation by computing SPICE's accuracy for both ICA and TCA tasks (\textit{SWE-L-Based Accuracy Evaluation}). Second, we conduct a \textit{manual evaluation}, where two labelers independently assess whether SPICE labels are correct and whether the provided rationale is appropriate. In cases of disagreement, a third labeler adjudicates to determine the final label. Finally, to better understand the quality of SPICE-generated labels, we conduct an experiment in which we determine whether there is a relationship between the number of top scaffolds that solved a given SWE-V instance and the ICA and TCA labels produced by SPICE (Scaffold-based Evaluation). We detail our evaluation setup below. 

\smallskip\noindent\textit{SWE-L-based Accuracy Evaluation}:

\tinyskip\noindent\textit{-- Evaluation Set.} We sampled 110 evaluation instances using quota sampling, selecting 10 instances from each of the 12 projects in the \textsc{SWE-L} dataset~\cite{baltes2022sampling}. This strategy enables us to assess the generalizability of SPICE across a diverse range of projects on both ICA and TCA tasks. For two projects with fewer than 10 available instances (\textit{flask} had only one instance and \textit{seaborn} had nine), we selected all available instances, resulting in a final total of 110 instances. 
While 110 instances may appear small at first glance, it exceeds the 91 instances that would result from randomly sampling \textsc{SWE-L} with a 95\% confidence level and a 10\% error margin (a common practice in Empirical SE ~\cite{bhatia2020study}). Given these factors, we consider our 110-instance evaluation set to be both robust and justified for this study.

For these instances, we did not use the final ICA\textsubscript{SWE-L-label} and TCA\textsubscript{SWE-L-label} provided by SWE-L moderators. Instead, we derived reference labels via majority vote across the individual ICA and TCA annotations from SWE-L labelers. This approach mitigates potential moderator bias~\cite{carletta1996assessing,cohen1960coefficient} and ensures a fairer comparison, as SPICE also uses majority voting to determine its ICA and TCA labels.

\tinyskip\noindent\textit{-- FM Selection.} To ensure a representative evaluation, we selected a diverse mix of open- and closed-source FMs, including both reasoning-oriented and general-purpose models. We include open-source models like DeepSeek R1 Reasoner~\cite{deepseek_reasoning_model} and closed-source models from OpenAI~\cite{openai_models}: GPT-4o, GPT‑4o mini, and GPT‑4.1. We also evaluate locally hosted models using the Ollama framework~\cite{ollama_library}, including DeepSeek 32B (Ollama-Deepseek-R1:32b), Qwen‑14B (Ollama-Qwen2.5:14b), and Qwen‑7B (Ollama-Qwen2.5:7b). This setup spans a range of parameter sizes and deployment contexts (covering both locally hosted and externally hosted environments) allowing us to comprehensively assess SPICE's performance across varied FM capabilities.

Finally, we ran SPICE on all 110 evaluation instances, computed its ICA\textsubscript{SPICE-label} and TCA\textsubscript{SPICE-label}, and measured their accuracy against the corresponding \textsc{SWE-L} reference labels. Note that we use SWE-L instead of SWE-V since SWE-V only has positive samples (i.e., instances with well-specified issues and adequate test coverage).

\smallskip\noindent\textit{Manual Agreement-based Evaluation}:

\noindent~We conducted a complementary manual evaluation to directly assess the validity of SPICE generated labels using the best performing FM from the previous step and their associated rationales. As noted in Section~\ref{sec:pre-study}, we observed low IRA among labelers in the \textsc{SWE-L} dataset. Therefore, while SPICE may exhibit high accuracy on the \textsc{SWE-L} evaluation, such scores may not fully reflect its real-world usefulness. In this setup, labelers reviewed the ICA\textsubscript{SPICE-label} and TCA\textsubscript{SPICE-label} along with their corresponding rationales, and responded to a simple but structured questionnaire.

\begin{tcolorbox}[colback=gray!10, colframe=gray!50, boxrule=0.3pt, arc=2pt, left=4pt, right=4pt, top=2pt, bottom=2pt]
\begin{small}
\begin{enumerate}[wide = 0pt]
    \item Do you ``agree'' with the SPICE tool's final label on \textbf{issue clarity}? [y/n]
    \item Give reason for the above.
    \item Do you ``agree'' with the SPICE tool's final label on \textbf{test adequacy}? [y/n]
    \item Give reason for the above.
\end{enumerate}
\end{small}
\end{tcolorbox}

Four labelers from our annotation team participated in evaluating the correctness of SPICE by reviewing the TCA\textsubscript{SPICE-label} and ICA\textsubscript{SPICE-label}. Each SPICE-generated label was independently evaluated by two of the four labelers. This validation effort spanned two days, with 24 instances annotated per day (48 instances in total). In cases of disagreement, a third labeler adjudicated by reviewing the ICA\textsubscript{SPICE-label} and TCA\textsubscript{SPICE-label}, their corresponding rationales, and the questionnaire responses from both initial labelers. Finally, we computed the accuracy of the ICA\textsubscript{SPICE-label} and TCA\textsubscript{SPICE-label} by comparing them with the final adjudicated labels derived from the questionnaire.

\smallskip\noindent\textit{Scaffold-based Evaluation}: To get further insights into the quality of the labels produced by SPICE, we carry out the following experiment. We first visit the SWE-V leaderboard\cite{swebench2025leaderboard} and identify the top-25 scaffolds (we tick the ``Checked'' filter). Next, we clone the SWE-Bench experiments repository (https://github.com/swe-bench/experiments) and visit the \texttt{main/evaluation/verified path}. This path contains several subdirectories, each representing the prediction results of a scaffold shown in the leaderboard webpage. After manually mapping the names displayed in the leaderboard to the subfolder names, we determine the list of ``resolved'' instances for each scaffold. Next, we compute a metric called resolution rate for each SWE-V instance, which we define as the fraction of scaffolds that resolved that instance (x / 25). We now run SPICE for all SWE-V instances to compute their ICA and TCA labels. Our hypothesis is that instances that are well-specified and have adequate test coverage will have a higher resolution rate than those that are poorly-specified and have inadequate test coverage. Hence, we test whether ICA and TCA have an effect on the resolution rate using Anova tests (\textit{alpha} = 0.05). We used GPT-4.1 to perform this experiment on 250 random SWE-V instances.

\smallskip\noindent\textbf{Results.} \textbf{(SWE-L-based Accuracy Evaluation) SPICE\textsubscript{G-4o-m} achieves an accuracy of 87.3\% on ICA, while SPICE\textsubscript{DS-R} achieves 68.5\% on TCA.} We interpret these scores as lower bounds for correctness, since the cases in which SPICE and SWE-L agree with each other are likely to represent ``correct'' labels more often than not. Hence, we consider our ICA and TCA accuracies to represent solid lower bounds for label correctness (especially considering the pioneering nature of SPICE).

Table~\ref{tab:rq1_different_model_performances} summarizes SPICE's performance across all studied FMs on both ICA and TCA tasks. As shown, SPICE consistently performs better on ICA than on TCA. Even the lowest-performing model SPICE\textsubscript{Q-7} (which we were unable to evaluate on TCA due to segmentation faults in Aider) achieves an accuracy of 68.2\% on ICA, which matches our best TCA score. We hypothesize that this discrepancy arises from the inherent complexity of the TCA task. Accurate TCA labeling requires SPICE to process and reason over significantly more contextual information, as discussed in Section~\ref{sec:tool}, which likely exceeds the capabilities of smaller models.

\smallskip\noindent\textbf{(Manual Agreement-based Evaluation) SPICE correctly labeled ICA 93.5\% of the time, and TCA 60\% of the time.} These results suggest that SPICE can be reliably adopted for ICA, but that TCA\textsubscript{SPICE-label} should be used with caution and accompanied by human review.

That said, we identified promising mitigation signals. Specifically, when evaluators \textit{disagreed} with TCA\textsubscript{SPICE-label}, the inter-rater agreement (IRA) among SPICE's three rationale generations was very low (IRA = -0.059, i.e., below chance). In contrast, when evaluators \textit{agreed} with the label, the corresponding rationale agreement was substantially higher (IRA = 0.45). Users can leverage rationale consistency across SPICE runs as a confidence indicator when interpreting TCA predictions~\cite{zhang2025d3}.

While SPICE's performance on TCA is imperfect, we believe the trade-off is reasonable, especially given the task's inherent difficulty and the fact that SPICE is the first automated tool tackling it. Moreover, our manual evaluation revealed a systematic trend: SPICE tends to be overly conservative, occasionally marking simple but valid test cases as inadequate. This ``over-strictness'' is a known limitation that can affect TCA\textsubscript{SPICE-label} reliability.

However, this does not negate SPICE's usefulness. Our evaluators reported that even when TCA\textsubscript{SPICE-label} was incorrect, reviewing the associated rationales allowed them to quickly identify and correct the error. This aligns with findings in cognitive psychology showing that comparisons, even with incorrect explanations, can help humans make better judgments than when reasoning from scratch~\cite{gligoric2024can}. Nonetheless, we acknowledge that SPICE's TCA capability remains an area for future improvement.

\begin{table}[ht]
  \centering
  \scriptsize
  \caption{Accuracy (\%) for issue and test labeling.}
  \label{tab:rq1_different_model_performances}
  \begin{tabular*}{\columnwidth}{@{\extracolsep{\fill}}lrr}
    \toprule
    \textbf{FM} & \textbf{ICA Accuracy (\%)} & \textbf{TCA Accuracy (\%)} \\
    \midrule
    \textbf{GPT-4o-mini (G-4o-m)} & \textbf{87.3} & 41.8 \\
    GPT-4.1 (G-4.1)                & 82.7          & 65.5 \\
    GPT-4o (G-4o)                  & 81.2          & 68.2 \\
    DeepSeek Reasoner (DS-R)      & 79.6          & \textbf{68.5} \\
    DeepSeek 32B (DS-32)\textsuperscript{*} & 76.8 & 43.2 \\
    Qwen 14B (Q-14)\textsuperscript{*}     & 75.5 & -- \\
    Qwen 7B (Q-7)\textsuperscript{*}       & 68.2 & -- \\
    \bottomrule
  \end{tabular*}
  \vspace{0.5em}
  \scriptsize \textsuperscript{*}Locally hosted models, served by Ollama, are 4-bit quantized.
\end{table}

\begin{figure}[!htbp]
  \centering
  \includegraphics[width=0.9\columnwidth,keepaspectratio]{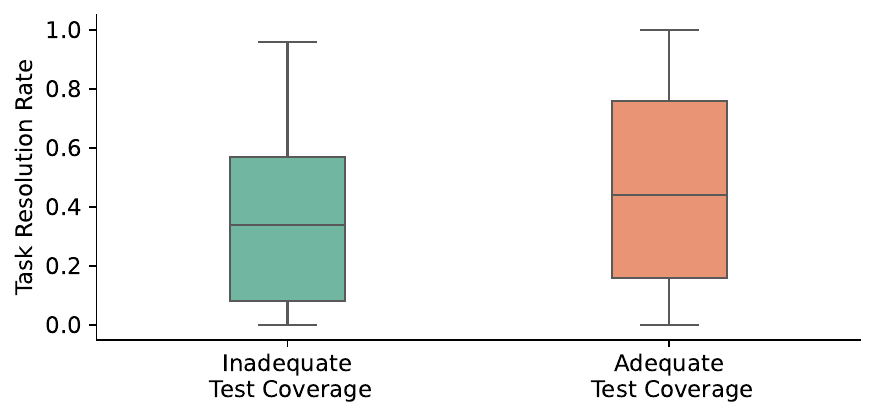}
  \caption{A comparison between the task resolution rate for SWE-V instances with inadequate (green) and adequate (orange) test coverage.}
  \label{fig:r1_good_tests_boxplot}
\end{figure}

\smallskip\noindent\textbf{(Scaffold-based Evaluation)
There is a statistically significant main effect of TCA labels on the resolution rate.} We first analyze the effect of TCA on the resolution rate. Instances with adequate test coverage have a higher resolution rate (n = 198, median = 0.44) compared to those with an inadequate test coverage (n = 52, median = 0.32). This difference is further supported by a one-way ANOVA, which indicates a statistically significant effect of the TCA label on the task resolution rate (F = 6.3, p = 0.01), reinforcing the reliability of SPICE labels. The distributions can be seen in Figure~\ref{fig:r1_good_tests_boxplot}. 

Next we analyze ICA. The SPICE labels indicate that 241 instances were classified as well-specified, while only 9 were classified as underspecified. Since all instances in SWE-V are positive, SPICE achieves high ICA accuracy by closely mirroring this distribution. Because the ICA labels are highly imbalanced (only 9 underspecified cases), assumptions of ANOVA regarding group size and variance homogeneity are difficult to meet, making inference unreliable. In particular, only three instances are both underspecified and have inadequate test coverage, rendering a two-way ANOVA also difficult to interpret. Notably, the lower TCA accuracy (Table 1) introduces greater diversity across labels, which is beneficial for evaluating task resolution rates on the SWE-V leaderboard.\label{subsec.RQ1}
\subsection{\rqtwo}

\noindent\textbf{Motivation.} 
Although Section~\ref{sec:pre-study} shows that \textsc{SWE-L} labels suffer from low IRA, the SWE-V benchmark remains popular because those labels are accompanied by a human-written rationale. These rationales make post-hoc verification and debugging feasible, which is critical for a dataset that is intrinsically hard to label. For SPICE to gain similar trust and adoption, its FM-generated rationales must approach the quality of those written by \textsc{SWE-L} labelers.  We therefore measure how closely SPICE rationales resemble human rationales.

\tinyskip
\noindent\textbf{Approach.} To answer this RQ, we use the same 110 instances and assess the semantic similarity between the rationales written by Human labelers and SPICE labelers. We treat each of the three runs in SPICE as equivalent to a SWE-L labeler and cover rationales for both ICA and TCA. To enable consistent comparison across annotations, we abstract SWE-L annotators into three rotating ``\textit{seats}'' (Human L1-L3), representing the three annotator roles per instance. These seats are filled by different subsets of the 93 human annotators in SWE-L, allowing us to compute semantic similarity without identity mismatches across instances (see Section~\ref{sec:background}).

To compute semantic similarity, we embed the three SWE-L rationales per instance into a single document using the \texttt{text-embedding-3-large}~\cite{embedding_model} embedding model. We do this both individually and with all three human rationales combined. We apply the same embedding procedure to SPICE rationales. Next, we compute the cosine similarity between the resulting embeddings (humans vs SPICE) for comparison. To ensure that SPICE-produced rationales are neither overly complex nor simplistic, we compare the word counts of SPICE and SWE-L rationales. For this RQ, we use SPICE\textsubscript{G-4o-m} for ICA and SPICE\textsubscript{DS-R} for TCA, as they demonstrated the best performance in the previous RQ.





\begin{figure}[!ht]
  \centering
  \includegraphics[width=0.8\columnwidth,keepaspectratio]{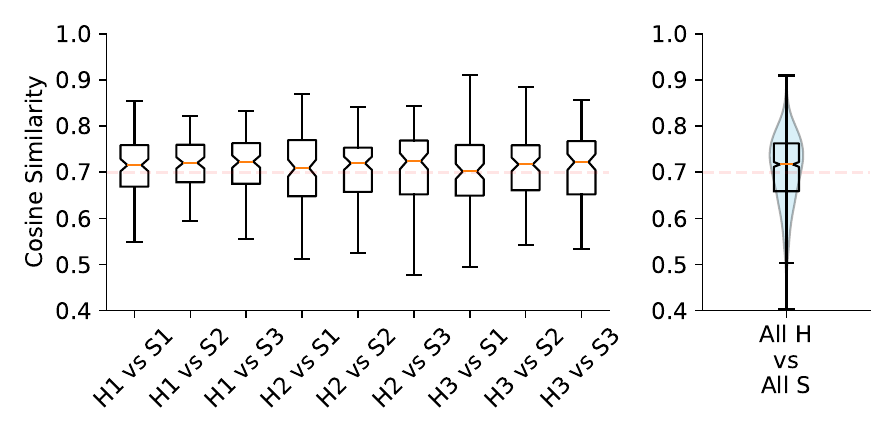}
  \caption{Results for ICA. (Left) shows semantic similarity between each Human labeler and SPICE Labeler. (Right) shows the semantic similarity between all Human rationales corpus and all SPICE-generated rationales corpus.}
  \label{fig:rq2_issue_similarities}
\end{figure}

\begin{figure}[!ht]
  \centering
  \includegraphics[width=0.8\linewidth,keepaspectratio]{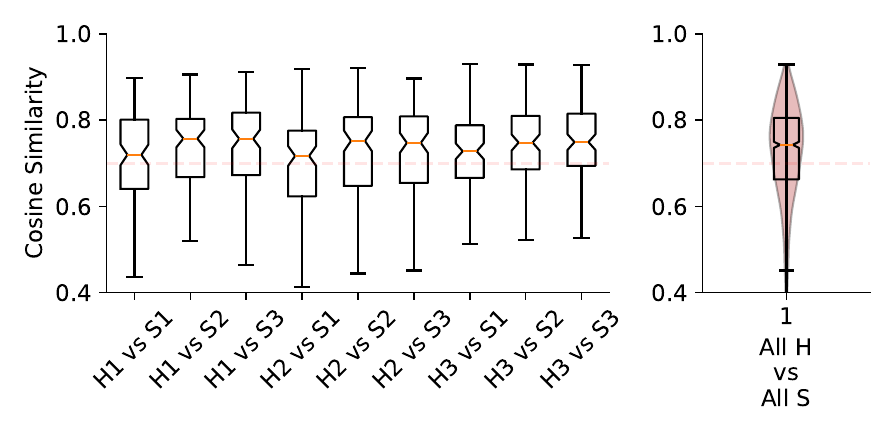}
  \caption{Results for TCA. Explanation analogous to Figure 2.}
  \label{fig:rq2_test_similarities}
\end{figure}

\tinyskip \noindent\textbf{Results.} \textbf{SPICE rationales exhibit strong semantic alignment with human rationales, achieving a median similarity of 0.72 for ICA and 0.743 for TCA (Figure~\ref{fig:rq2_issue_similarities}, \ref{fig:rq2_test_similarities}).} These results suggest that SPICE labelers consistently produce explanations that reflect human reasoning.

In ICA, the corpus-level median similarity between SPICE and human rationales is 0.72 (Figure~\ref{fig:rq2_issue_similarities}, right), and individual comparisons across the three seats yield values ranging from 0.71-0.72 (Figure~\ref{fig:rq2_issue_similarities}, left).

\begin{figure}[!htbp]
  \centering
  \includegraphics[width=0.9\columnwidth,keepaspectratio]{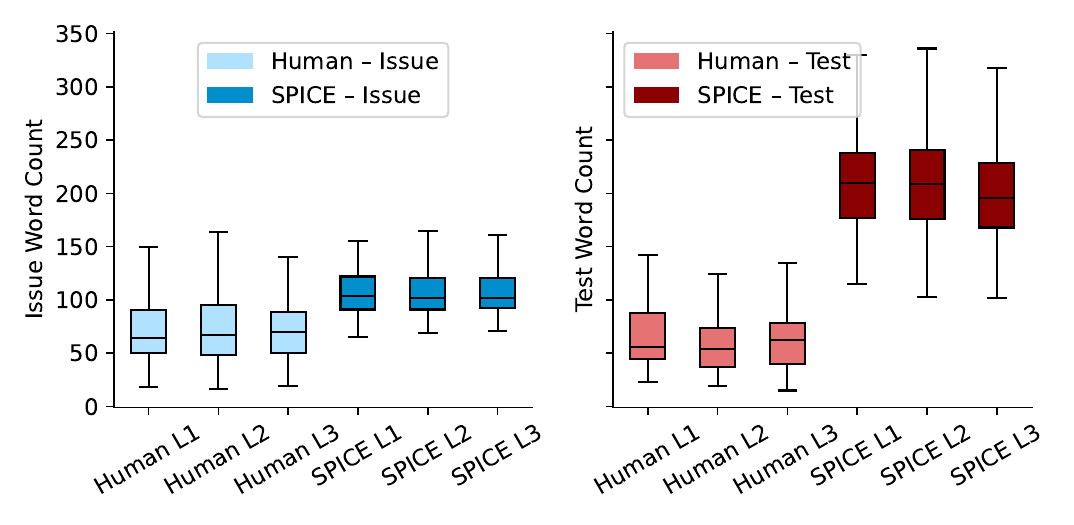}
  \caption{Issue labeling (left) and test labeling (right) rationale word count comparison for Human Labelers from SWE-L and SPICE labelers.}
  \label{fig:rq2_wordcount}
\end{figure}

\tinyskip\noindent\textbf{SPICE rationales are also longer.} The median word count is 102 compared to 67 for human annotators (Figure~\ref{fig:rq2_wordcount}, left), providing roughly 1.5x more explanation for ICA. 

For TCA rationales, SPICE again shows strong semantic overlap (median similarity of 0.743; Figure~\ref{fig:rq2_test_similarities}), and the explanations are significantly more detailed: 420 words on average versus just 58 for human annotators (Figure~\ref{fig:rq2_wordcount}, right), or nearly 7x longer. This verbosity stems from the structured prompt used in the TCA pipeline, which instructs the model to provide a rationale for its answer as well as a counter-example (e.g., some valid patch that would be missed by the tests) when it chooses a score that is higher than zero.

\label{subsec.RQ2}
\subsection{\rqthree}
\label{sec:three}

\noindent\textbf{Motivation.} 
As FM-based labeling becomes viable for large-scale SE datasets (e.g., the 21.5k-instance SWE-bench and its multi-thousand-instance variants), practitioners face a crucial \textit{quality vs cost} dilemma. While SPICE unlocks high-throughput annotation, each inference incurs real monetary and time expenses. Quantifying these expenses across model choices, token volumes, and latency profiles enables informed decisions about which FM configurations balance accuracy with budget and turnaround requirements. This analysis provides actionable guidance to users of our tool for efficiently labeling SWE-bench-like data at scale, while avoiding unexpected financial or time overruns.

\noindent\textbf{Approach.}
To frame our cost-and-time analysis for this research question, we decompose the evaluation into two complementary components. 
First, we capture SPICE's own operational parameters, including token usage and inference latencies, using our logging infrastructure. Then, we translate those metrics into API pricing estimates based on public rate cards. Both parts are explained in detail below:

\tinyskip\noindent\textit{-- Parameter Estimation for SPICE.} Our SPICE tool leverages  \texttt{litellm}~\cite{litellm_docs} as a dependency and avails all the benefits of quick logging. We obtain the issue logs, along with the Aider logs for test logging, and obtain both input and output tokens (the distinction is important because of different pricing), and the time taken for inference calls along with the internal tool working (e.g., querying the relevant files by Aider). For all estimates, like input and output tokens, we used SPICE\textsubscript{G-4o-m} for ICA and SPICE\textsubscript{DS-R} for TCA (i.e., we use the same 110 instances from RQ1 and RQ2). While estimating time, we did not use Deepseek API, since Deepseek's inference can fluctuate based on server load~\cite{deepseekTimeout2025,deepseekGitHub2025,byteplusInfra2025,milvusLatency2025}; preventing us from getting an accurate estimate. Hence, we used the time logs from queries to OpenAI API using SPICE\textsubscript{G-4o-m} for both ICA and TCA for estimating time.

\tinyskip\noindent\textit{-- Estimation of API pricing.} We obtain input and output cost values from public pricing pages~\citep{openaipricing,deepseekpricing}. While server‑side caching may reduce effective cost, it cannot be computed/estimated at the client side, and not all FM providers provide this, we exclude it from our estimates, and report the worst case scenarios.

\begin{figure}[!ht]
  \centering
  \includegraphics[width=\columnwidth,keepaspectratio]{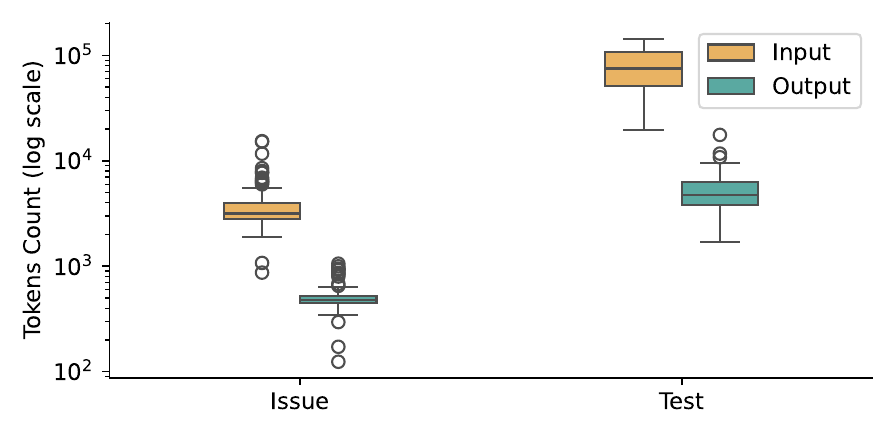}
  \caption{Distribution of input tokens (sent to the API), and output tokens (received from the API) for SPICE\textsubscript{G-4o-m} for ICA and SPICE\textsubscript{DS-R} for TCA.} 
  \label{fig.rq3_token_distribution}
\end{figure}

\tinyskip\noindent\textbf{Results.}
\textbf{SPICE sends a notably high amount (median 75,000) of input tokens (i.e., prompt tokens) for TCA and receives a median of 4,700 output tokens (i.e., model response tokens).}
Figure~\ref{fig.rq3_token_distribution} illustrates the token distribution for issue and test labeling.  The high number of output tokens for TCA arises because SPICE leverages additional contextual information from multiple related files associated with each patch. In contrast, \textbf{ICA labeling requires significantly fewer tokens, with a median of only 3,162 output tokens per instance and a median of 475 input tokens per instance.} This is also reflected in the time taken for generating issue labels compared to test labels.
Labeling one issue takes a median time of 10 seconds, whereas labeling one test is approximately 8 times slower at 83 seconds. Naturally, the actual time is dependent on server load, network bandwidth, etc., and can vary. 



Table~\ref{tab:model-costs} provides the estimated labeling costs for 1,000 instances based on median token usage, should these models be used for both issue and test labeling. These costs combine total input and output tokens with their respective pricing structures. As per our default setting (SPICE\textsubscript{G-4o-m} for ICA and SPICE\textsubscript{DS-R} for TCA),  labeling 1,000 instances using SPICE costs approximately \$5.1 (\$0.11 for issues and \$4.99 for tests). 


\begin{table}[ht]
  \centering
  \scriptsize
  \caption{API Cost estimates for labeling 1K instances using the median of 9,486 input tokens and 4,700 output tokens for both test and issue labeling combined.}
  \label{tab:model-costs}
  \begin{tabular*}{\columnwidth}{@{\extracolsep{\fill}}lrrr}
    \toprule
    Model            & In.\ Cost (\$) & Out.\ Cost (\$) & Total (\$) \\
    \midrule
    GPT-4o (G-4o)           & 7.91 & 47.00 & 54.91 \\
    GPT-4.1 (G-4.1)           & 6.32 & 37.60 & 43.92 \\
    DeepSeek Reasoner (DS-R) & 1.74 &  2.59 &  4.32 \\
    GPT-4o-mini (G-4o-m)       & 0.47 &  2.82 &  3.29 \\
    \bottomrule
  \end{tabular*}
\end{table}

Additional measures can be taken for reducing costs. From our experiments, we observe that offline quantized models performed reasonably well in ICA. For instance, SPICE\textsubscript{Q-14} achieves approximately 75\% accuracy in ICA, which is only 12 percentage points below the top-performing model (Table~\ref{tab:rq1_different_model_performances}). Additional cost savings can be achieved by modifying SPICE's default parameters, such as the number of SPICE labelers, assigning different models to different SPICE labelers, or limiting the size of the code context retrieved by Aider (which SPICE conveniently supports via configurable settings). The users of our tool may wish to evaluate the cost/performance trade-offs by tinkering around with these parameters.
\label{subsec.RQ3}

\section{The SPICE\spice-Bench Dataset}

A core contribution of this paper is the release of SPICE-Bench. SPICE-Bench comprises 6,802 SPICE-labeled instances (code-repair tasks) collected from 291 repositories sampled from the SWE-Gym dataset\cite{pan2024training}. This number of repositories is much larger than that of SWE-V (12), making SPICE-Bench significantly more diverse and heterogeneous.

58.5\% (3,981) of the instances in SPICE-Bench contain a well-specified issue, while 49.2\% (3,342) contain adequate test coverage. 31.8\% (2,164) of instances contain both a well-specified issue and adequate test coverage. This set of 2,164 ``good'' instances is 4.3x the size of SWE-V (500).

Analogously to SWE-L, SPICE-Bench also includes labels for difficulty (0 indicates > 15 mins of effort, 1 indicates 15 mins-1hr, 2 indicates 1-4 hrs, and 3 indicates 4+ hrs.). Out of the 2,164 good instances, 7.8\% (169) scored 0, 27.8\% (602) scored 1, 60.9\% (1,218) scored 2, and 3.5\% (75) scored 3 for difficulty.

More information about SPICE\spice-Bench is available on Hugging Face\textsuperscript{*}\footnote{\textsuperscript{*}The URL will be made available after acceptance.}. 

\section{Related Work}
\label{sec.related}

\noindent \textbf{Labeling Automation.} In the following, we discuss labeling automation both in a general context and in the SE context.

\tinyskip \noindent \textit{-- General context.} The advent of FMs, and particularly LLMs, in 2020~\cite{brown2020language} led to a transformation of the labeling landscape: the cumbersome traditional ML infrastructure (e.g., random forest classifier plus task-specific data curation) started to give room to end-to-end FM-based solutions. Research studies reported FM labeling efforts across a plethora of domains and applications, including: sentiment, topic, and content segmentation on multilingual text~\cite{rathje2024gpt}, content moderation in online communities~\cite{kolla2024llm}, labeling sections of contracts or laws~\cite{savelka2023unreasonable}, and annotation of breast cancer pathology reports~\cite{sushil2024comparative}.

\tinyskip \noindent \textit{-- Software Engineering context.} Labeling code artifacts, such as commits and code comments, have traditionally required significant manual effort or bespoke ML pipelines (e.g., ~\cite{yu2019improving}). Recently, researchers have explored the potential of FMs to automate the labeling across several domains and applications. For instance, ~\citet{sazid2024commit} evaluated GPT-3 for commit messages labeling into maintenance categories (corrective, adaptive, perfective), achieving 75.7\% accuracy in zero/few-shot settings, on par with a SOTA supervised ML classifiers. \citet{sheikhaei2024empirical} compared in-context learning and finetuning FMs performance for labeling Self-Admitted Technical Debt (SATD) in code comments. The authors found that finetuned FM outperformed even the largest FM used in their study by 9.2\%. ~\citet{lambert2024identification} leveraged FMs for SATD identification by exploring the performance of three FMs across several in-context learning strategies. FMs also support other SE tasks like bug-report triaging by labeling severity, priority and duplication, thus guiding the bug-fix order. ~\citet{arshad2024sevpredict} fine-tuned an FM to label bug severity into ``blocker'', ``critical'', and ``minor''. For labeling duplicate bug reports, ~\citet{zhang2023cupid} developed CUPID, which leverages FMs to extract keywords from bug descriptions to retrieve similar past reports, thereby outperforming existing de-duplication methods. In a more visionary paper, \citet{Gerosa2024} explore the potential of FMs as substitutes or supplements for human participants in qualitative software engineering research, demonstrating FMs' abilities to emulate interviews, focus groups, surveys, and observational studies. While highlighting promising applications, the authors underscore the need for caution due to ethical, fairness, and methodological challenges.

\tinyskip \noindent \textbf{SWE-bench Limitations and Variants.} While SWE-V largely superseded SWE-bench, it still has clear limitations. The dataset is relatively small (500 instances), all 12 projects are written in Python, and empirical evidence suggests that newer models were trained with SWE-bench data and have memorized issue-fix pairs~\cite{zhang2025swebenchgoeslive} (SPICE can be used to create fresh datasets and mitigate the risks of benchmark data leakage during pre-training). Other research studies have recently flagged additional data quality issues in SWE-V~\cite{aleithan2024swebenchenhancedcodingbenchmark}. For instance, \citet{wang2025solved} indicate that SWE-V patches do not cover all corner cases and incorrect patches can still sneak through. The authors call for more robust evaluation practices, including running all available tests, filtering incorrect patches, and strengthening benchmarks with differentiating tests generated by their proposed tool (PatchDiff) to build a sustainable patch validation ecosystem.


As a response to the aforementioned limitations, several SWE-bench-like datasets have been proposed. For example, in terms of broadening the scope of SWE-V, ByteDance's SWE‑bench-Multilingual~\cite{zan2025multi} includes 300 tasks in nine languages and SWE‑bench Multimodal~\cite{yang2024swe} include approximately 600 GUI-based JavaScript tasks with screenshots. 
A non-exhaustive list of SWE-V variants include: SWE-Bench+~\citep{aleithan2024swebenchenhancedcodingbenchmark}, SWE‑Lancer~\cite{miserendino2025swe}, SWE‑Gym~\cite{pan2024training}, SWE-bench-Live~\citep{zhang2025swebenchgoeslive}, and SWE-bench Lite.

\tinyskip \noindent \textbf{The role of SWE-bench-like datasets in model training.} In addition to supporting the evaluation of automated program repair techniques (e.g.,  OpenHands~\cite{openhands}), SWE-bench-like datasets also have a fundamental role in supporting model training. For example, SkyRL~\citep{cao2025skyrl} is an open-source reinforcement learning (RL) training pipeline to enable FMs to function as real-world, long-horizon agents. Built on top of VeRL and OpenHands, SkyRL extends existing RL frameworks to support multi-turn tool use, asynchronous rollouts, and scalable environment execution, ultimately addressing challenges in stateful, dynamic tasks like those in SWE-V. The authors report that SkyRL achieves significant performance improvements across various model sizes using minimal training data, demonstrating its efficiency and effectiveness in complex environments.  As another key example, \citet{pan2024training} report that training FM-powered agents in SWE-Gym led to up to 19\% improvement in task resolution on the SWE-V and SWE-bench lite benchmarks. Further performance gains on both benchmarks were achieved by using verifier models during inference.

\section{Threats to Validity} \label{sec.threats}

To ensure the credibility and robustness of our findings, we systematically identify potential threats to validity and describe strategies implemented to mitigate them during the development and evaluation of the SPICE tool.

\noindent\textbf{Construct Validity}. Our labeling scale, adapted from the SWE-V dataset, employs an ordinal scoring system ranging from 0 to 3, assuming equidistant intervals between categories. These ordinal scores are subsequently converted into binary categories (i.e., good issue/test = 0 or 1). Instances receiving higher scores (2 or 3) are excluded from the selection process. A potential threat arises because real-world constructs, such as \textit{issue clarity} or \textit{test coverage}, may not exhibit linear progressions. As a consequence, distinguishing between boundary conditions (i.e., 1 (keep) and 2 (discard)) can be inherently challenging. However, this threat is minimized due to the rigorous heuristic-based labeling approach adopted by OpenAI in the SWE-V dataset, involving careful curation by professional teams involved in the rubric curation at OpenAI. Another construct threat is associated with prompt wording bias. The phrasing of prompts inherently influences model judgments related to clarity, adequacy, and perceived effort. To address this, we publish the specific prompts used and encourage replication studies within the research community to validate our findings independently.

\noindent\textbf{Internal Validity.} 
Outputs generated by FMs inherently vary due to stochastic factors such as temperature settings and random seeds. This variability poses a risk similar to human annotator inconsistencies. To mitigate this risk, we employ a multi-labeler voting strategy, effectively minimizing bias introduced by any single annotator or instance of stochasticity.\\ 
In addition, although SPICE can estimate task difficulty, we do not evaluate it in our research questions, because our primary objective is selecting solvable issues rather than measuring annotator effort. Moreover, testing is costly. Each additional experiment adds non‐trivial monetary expense.

However, in the working of our tool, since difficulty‐estimation procedure is exactly the same as TCA's except for a modified prompt, we assume that our RQ2 and RQ3 results will be similar to that of TCA's. Specifically, SPICE difficulty rationales in RQ2 should exhibit the similar (7x) increase in length, and that token usage and cost projections in RQ3 should be comparable. Although the evaluation of  our tool's full capabilities remains a threat, it does not diminish our tool's value in filtering of ``unsolvable'' issues. Moreover, difficulty labels are provided in our SPICE-bench release, along with the evaluation prompt to support future work.

\noindent\textbf{External Validity.} In our RQs, we used SPICE to label issues and tests from SWE-V (python projects). Therefore, the performance of SPICE on different datasets (e.g., involving other programming languages or issues written using unusual conventions) has not been assessed and thus pose a threat to external validity. In any case, to mitigate the threat, SPICE incorporates flexible prompt management through easily configurable and version-controlled files, allowing easy adjustments to accommodate \textit{evolving} data characteristics (e.g., data drift). 
Another threat arises from the selection of Aider's context size. In projects characterized by low cohesion (how closely components are related internally) and high coupling (dependency between modules), much more context might be required than Aider's default context selection quota. However, this threat is significantly mitigated by exposing this parameter as an easily adjustable user setting in SPICE. Finally, our estimated annotation cost of \$0.51 per instance is an upper bound, and server-side caching should reduce this amount, as the model sees the same prompt three times. Moreover, this cost is contingent on current API pricing structures and may vary significantly depending on the choice and configuration of FMs, as well as organizational factors external to the research environment.



\section{Conclusion}\label{sec.conclusion}

In this work, we introduced SPICE\spice, an automated labeling pipeline for SWE-bench-like datasets that generates interpretable, rubric-aligned labels for issue clarity, test coverage, and effort estimation. Motivated by the high cost, low replicability, and limited scalability of manual annotation (Section~\ref{sec:pre-study}), we designed SPICE to synthesize high-quality labels using prompt engineering, context-aware analysis powered by Aider, and multi-run consensus. Our results demonstrate SPICE's effectiveness in reproducing expert labels at a fraction of the cost, enabling scalable dataset construction for evaluating and training software-focused FMs. 

To support the community, we publicly release both the \textbf{SPICE\spice  tool} and \textbf{SPICE\spice-bench}. SPICE-bench is over \textbf{13 times larger} than SWE-V and is intended to accelerate FM research by enabling fine-tuning, benchmarking, and tool development at scale.

Looking ahead, SPICE's potential is promising. While the current version remains in an early alpha stage, ongoing experimentation with different model configurations and prompting strategies suggests substantial headroom for improvement. Moreover, SPICE is built atop two fast-evolving components: FMs and Aider. In particular, Aider benefits from an active open-source community, which continues to enhance its capabilities for code navigation and context retrieval. As these underlying technologies mature, we expect SPICE to become increasingly more accurate, efficient, and adaptable for large-scale SE data curation.

\section*{Acknowledgments}
Any opinions, findings, and conclusions, or recommendations expressed in this material are those of the author(s) and do not reflect the views of Huawei. In addition, we used AI to polish the language of certain sections. All outputs were manually reviewed and verified in compliance with ACM/IEEE generative AI authorship guidelines.

We would like to acknowledge the following people’s contributions to this work (ordered by last name): Yuan Guowen, Hu Jiajun, and Chong Chun Yong.

\bibliographystyle{IEEEtranN}
\begin{small}
\bibliography{references}

\begin{thebibliography}{58}
\providecommand{\natexlab}[1]{#1}
\providecommand{\url}[1]{#1}
\csname url@samestyle\endcsname
\providecommand{\newblock}{\relax}
\providecommand{\bibinfo}[2]{#2}
\providecommand{\BIBentrySTDinterwordspacing}{\spaceskip=0pt\relax}
\providecommand{\BIBentryALTinterwordstretchfactor}{4}
\providecommand{\BIBentryALTinterwordspacing}{\spaceskip=\fontdimen2\font plus
\BIBentryALTinterwordstretchfactor\fontdimen3\font minus \fontdimen4\font\relax}
\providecommand{\BIBforeignlanguage}[2]{{%
\expandafter\ifx\csname l@#1\endcsname\relax
\typeout{** WARNING: IEEEtranN.bst: No hyphenation pattern has been}%
\typeout{** loaded for the language `#1'. Using the pattern for}%
\typeout{** the default language instead.}%
\else
\language=\csname l@#1\endcsname
\fi
#2}}
\providecommand{\BIBdecl}{\relax}
\BIBdecl

\bibitem[OpenAI(2024)]{swebenchverified2024}
\BIBentryALTinterwordspacing
OpenAI, ``{Introducing SWE-bench Verified},'' 2024, accessed: 2025-05-04. [Online]. Available: \url{https://openai.com/index/introducing-swe-bench-verified/}
\BIBentrySTDinterwordspacing

\bibitem[Pan et~al.(2024)Pan, Wang, Neubig, Jaitly, Ji, Suhr, and Zhang]{pan2024training}
J.~Pan, X.~Wang, G.~Neubig, N.~Jaitly, H.~Ji, A.~Suhr, and Y.~Zhang, ``Training software engineering agents and verifiers with swe-gym,'' \emph{arXiv preprint arXiv:2412.21139}, 2024.

\bibitem[Zan et~al.(2025)Zan, Huang, Liu, Chen, Zhang, Xin, Chen, Liu, Zhong, Li, et~al.]{zan2025multi}
D.~Zan, Z.~Huang, W.~Liu, H.~Chen, L.~Zhang, S.~Xin, L.~Chen, Q.~Liu, X.~Zhong, A.~Li \emph{et~al.}, ``Multi-swe-bench: A multilingual benchmark for issue resolving,'' \emph{arXiv preprint arXiv:2504.02605}, 2025.

\bibitem[Yang et~al.(2024)Yang, Jimenez, Zhang, Lieret, Yang, Wu, Press, Muennighoff, Synnaeve, Narasimhan, et~al.]{yang2024swe}
J.~Yang, C.~E. Jimenez, A.~L. Zhang, K.~Lieret, J.~Yang, X.~Wu, O.~Press, N.~Muennighoff, G.~Synnaeve, K.~R. Narasimhan \emph{et~al.}, ``Swe-bench multimodal: Do ai systems generalize to visual software domains?'' \emph{arXiv preprint arXiv:2410.03859}, 2024.

\bibitem[swe(2025)]{swebench2025leaderboard}
``Swe-bench leaderboard,'' \url{https://www.swebench.com/}, 2025, accessed: 2025-05-15.

\bibitem[Cao et~al.(2025)Cao, Hegde, Li, Griggs, Liu, Tang, Pan, Wang, Malik, Hakhamaneshi, Liaw, Moritz, Zaharia, Gonzalez, and Stoica]{cao2025skyrl}
S.~Cao, S.~Hegde, D.~Li, T.~Griggs, S.~Liu, E.~Tang, J.~Pan, X.~Wang, A.~Malik, K.~Hakhamaneshi, R.~Liaw, P.~Moritz, M.~Zaharia, J.~E. Gonzalez, and I.~Stoica, ``Skyrl-v0: Train real-world long-horizon agents via reinforcement learning,'' \url{https://novasky-ai.notion.site/skyrl-v0}, 2025, accessed: 2025-05-15.

\bibitem[Dickson(2024)]{dickson2024code}
B.~Dickson, ``Code in pre-training data improves llm performance at non-coding tasks,'' https://venturebeat.com/ai/code-in-pre-training-data-improves-llms-performance-at-non-coding-tasks/, August 2024, accessed: 2025-05-15.

\bibitem[Zhang et~al.(2025{\natexlab{a}})Zhang, Chen, Ye, Yang, Chen, Wang, and Petzold]{zhang2025unveiling}
X.~Zhang, Z.~Z. Chen, X.~Ye, X.~Yang, L.~Chen, W.~Y. Wang, and L.~R. Petzold, ``Unveiling the impact of coding data instruction fine-tuning on large language models reasoning,'' in \emph{Proceedings of the AAAI Conference on Artificial Intelligence}, vol.~39, no.~24, 2025, pp. 25\,949--25\,957.

\bibitem[Carletta(1996)]{carletta1996assessing}
J.~Carletta, ``Assessing agreement on classification tasks: the kappa statistic,'' \emph{arXiv preprint cmp-lg/9602004}, 1996.

\bibitem[Cohen(1960)]{cohen1960coefficient}
J.~Cohen, ``A coefficient of agreement for nominal scales,'' \emph{Educational and psychological measurement}, vol.~20, no.~1, pp. 37--46, 1960.

\bibitem[Krippendorff(2011)]{krippendorff2011computing}
K.~Krippendorff, ``Computing krippendorff’s alpha-reliability,'' 2011.

\bibitem[Zhou et~al.(2023)Zhou, Zhu, Chen, Chen, Zhao, Chen, Lin, Wen, and Han]{zhou2023don}
K.~Zhou, Y.~Zhu, Z.~Chen, W.~Chen, W.~X. Zhao, X.~Chen, Y.~Lin, J.-R. Wen, and J.~Han, ``Don't make your llm an evaluation benchmark cheater,'' \emph{arXiv preprint arXiv:2311.01964}, 2023.

\bibitem[{Aider AI LLC}(2025)]{Aider2025}
{Aider AI LLC}, ``Aider: Ai pair programming in your terminal,'' \url{https://aider.chat/}, 2025, accessed: 2025-05-15.

\bibitem[Wang et~al.(2023)Wang, Wei, Schuurmans, Le, Chi, Narang, Chowdhery, and Zhou]{wang2023selfconsistencyimproveschainthought}
\BIBentryALTinterwordspacing
X.~Wang, J.~Wei, D.~Schuurmans, Q.~Le, E.~Chi, S.~Narang, A.~Chowdhery, and D.~Zhou, ``Self-consistency improves chain of thought reasoning in language models,'' 2023. [Online]. Available: \url{https://arxiv.org/abs/2203.11171}
\BIBentrySTDinterwordspacing

\bibitem[Jimenez et~al.(2023)Jimenez, Yang, Wettig, Yao, Pei, Press, and Narasimhan]{jimenez2023swe}
C.~E. Jimenez, J.~Yang, A.~Wettig, S.~Yao, K.~Pei, O.~Press, and K.~Narasimhan, ``Swe-bench: Can language models resolve real-world github issues?'' \emph{arXiv preprint arXiv:2310.06770}, 2023.

\bibitem[Chen et~al.(2021)Chen, Tworek, Jun, Yuan, de~Oliveira~Pinto, Kaplan, Edwards, Burda, Joseph, Brockman, Ray, Puri, Krueger, Petrov, Khlaaf, Sastry, Mishkin, Chan, Gray, Ryder, Pavlov, Power, Kaiser, Bavarian, Winter, Tillet, Such, Cummings, Plappert, Chantzis, Barnes, Herbert-Voss, Guss, Nichol, Paino, Tezak, Tang, Babuschkin, Balaji, Jain, Saunders, Hesse, Carr, Leike, Achiam, Misra, Morikawa, Radford, Knight, Brundage, Murati, Mayer, Welinder, McGrew, Amodei, McCandlish, Sutskever, and Zaremba]{humaneval}
M.~Chen, J.~Tworek, H.~Jun, Q.~Yuan, H.~P. de~Oliveira~Pinto, J.~Kaplan, H.~Edwards, Y.~Burda, N.~Joseph, G.~Brockman, A.~Ray, R.~Puri, G.~Krueger, M.~Petrov, H.~Khlaaf, G.~Sastry, P.~Mishkin, B.~Chan, S.~Gray, N.~Ryder, M.~Pavlov, A.~Power, L.~Kaiser, M.~Bavarian, C.~Winter, P.~Tillet, F.~P. Such, D.~Cummings, M.~Plappert, F.~Chantzis, E.~Barnes, A.~Herbert-Voss, W.~H. Guss, A.~Nichol, A.~Paino, N.~Tezak, J.~Tang, I.~Babuschkin, S.~Balaji, S.~Jain, W.~Saunders, C.~Hesse, A.~N. Carr, J.~Leike, J.~Achiam, V.~Misra, E.~Morikawa, A.~Radford, M.~Knight, M.~Brundage, M.~Murati, K.~Mayer, P.~Welinder, B.~McGrew, D.~Amodei, S.~McCandlish, I.~Sutskever, and W.~Zaremba, ``Evaluating large language models trained on code,'' 2021.

\bibitem[Austin et~al.(2021)Austin, Odena, Nye, Bosma, Michalewski, Dohan, Jiang, Cai, Terry, Le, and Sutton]{mbpp}
\BIBentryALTinterwordspacing
J.~Austin, A.~Odena, M.~Nye, M.~Bosma, H.~Michalewski, D.~Dohan, E.~Jiang, C.~Cai, M.~Terry, Q.~Le, and C.~Sutton, ``Program synthesis with large language models,'' 2021. [Online]. Available: \url{https://arxiv.org/abs/2108.07732}
\BIBentrySTDinterwordspacing

\bibitem[Aleithan et~al.(2024)Aleithan, Xue, Mohajer, Nnorom, Uddin, and Wang]{aleithan2024swebenchenhancedcodingbenchmark}
\BIBentryALTinterwordspacing
R.~Aleithan, H.~Xue, M.~M. Mohajer, E.~Nnorom, G.~Uddin, and S.~Wang, ``Swe-bench+: Enhanced coding benchmark for llms,'' 2024. [Online]. Available: \url{https://arxiv.org/abs/2410.06992}
\BIBentrySTDinterwordspacing

\bibitem[Urquhart(2013)]{urquhart2013grounded}
C.~Urquhart, \emph{Grounded Theory for Qualitative Research: A Practical Guide}.\hskip 1em plus 0.5em minus 0.4em\relax London: SAGE Publications, 2013.

\bibitem[Bhatia et~al.(2023)Bhatia, Eghan, Grichi, Cavanagh, Jiang, and Adams]{bhatia2023towards}
A.~Bhatia, E.~E. Eghan, M.~Grichi, W.~G. Cavanagh, Z.~M. Jiang, and B.~Adams, ``Towards a change taxonomy for machine learning pipelines: Empirical study of ml pipelines and forks related to academic publications,'' \emph{Empirical Software Engineering}, vol.~28, no.~3, p.~60, 2023.

\bibitem[Fabbri et~al.(2021)Fabbri, Kry{\'s}ci{\'n}ski, McCann, Xiong, Socher, and Radev]{fabbri2021summeval}
A.~R. Fabbri, W.~Kry{\'s}ci{\'n}ski, B.~McCann, C.~Xiong, R.~Socher, and D.~Radev, ``Summeval: Re-evaluating summarization evaluation,'' \emph{Transactions of the Association for Computational Linguistics}, vol.~9, pp. 391--409, 2021.

\bibitem[Krippendorff(2018)]{krippendorff2018content}
K.~Krippendorff, \emph{Content analysis: An introduction to its methodology}.\hskip 1em plus 0.5em minus 0.4em\relax Sage publications, 2018.

\bibitem[B{\"o}rstler et~al.(2023)B{\"o}rstler, Bennin, Hooshangi, Jeuring, Keuning, Kleiner, MacKellar, Duran, St{\"o}rrle, Toll, et~al.]{borstler2023developers}
J.~B{\"o}rstler, K.~E. Bennin, S.~Hooshangi, J.~Jeuring, H.~Keuning, C.~Kleiner, B.~MacKellar, R.~Duran, H.~St{\"o}rrle, D.~Toll \emph{et~al.}, ``Developers talking about code quality,'' \emph{Empirical Software Engineering}, vol.~28, no.~6, p. 128, 2023.

\bibitem[Bai et~al.(2024)Bai, Pei, Gu, Yang, and Ma]{bai2024special}
Y.~Bai, G.~Pei, J.~Gu, Y.~Yang, and X.~Ma, ``Special characters attack: Toward scalable training data extraction from large language models,'' \emph{arXiv preprint arXiv:2405.05990}, 2024.

\bibitem[Lin et~al.(2024)Lin, Dai, Verma, Ng, Jaillet, and Low]{lin2024prompt}
X.~Lin, Z.~Dai, A.~Verma, S.-K. Ng, P.~Jaillet, and B.~K.~H. Low, ``Prompt optimization with human feedback,'' \emph{arXiv preprint arXiv:2405.17346}, 2024.

\bibitem[{DeepSeek-AI and Team}(2025)]{deepseekr1}
\BIBentryALTinterwordspacing
{DeepSeek-AI and Team}, ``Deepseek-r1: Incentivizing reasoning capability in llms via reinforcement learning,'' 2025. [Online]. Available: \url{https://arxiv.org/abs/2501.12948}
\BIBentrySTDinterwordspacing

\bibitem[{Aider AI}(2023)]{AiderRepomap}
\BIBentryALTinterwordspacing
{Aider AI}, ``Repository map | aider,'' 2023, accessed: 2025-05-15. [Online]. Available: \url{https://aider.chat/docs/repomap.html}
\BIBentrySTDinterwordspacing

\bibitem[Baltes and Ralph(2022)]{baltes2022sampling}
S.~Baltes and P.~Ralph, ``Sampling in software engineering research: A critical review and guidelines,'' \emph{Empirical Software Engineering}, vol.~27, no.~4, p.~94, 2022.

\bibitem[Bhatia et~al.(2020)Bhatia, Wang, Asaduzzaman, and Hassan]{bhatia2020study}
A.~Bhatia, S.~Wang, M.~Asaduzzaman, and A.~E. Hassan, ``A study of bug management using the stack exchange question and answering platform,'' \emph{IEEE Transactions on Software Engineering}, vol.~48, no.~2, pp. 502--518, 2020.

\bibitem[{DeepSeek}()]{deepseek_reasoning_model}
{DeepSeek}, ``Reasoning model (deepseek-reasoner),'' \url{https://api-docs.deepseek.com/guides/reasoning_model}, accessed: 2025-05-15.

\bibitem[{OpenAI}()]{openai_models}
{OpenAI}, ``{Models -- OpenAI API},'' \url{https://platform.openai.com/docs/models}, accessed: 2025-05-15.

\bibitem[{Ollama Inc.}(2025)]{ollama_library}
{Ollama Inc.}, ``Ollama model library,'' \url{https://ollama.com/library}, 2025, accessed: 2025-05-15.

\bibitem[Zhang et~al.(2025{\natexlab{b}})Zhang, Zhang, Liu, Jin, Yang, Zheng, Liu, and Guo]{zhang2025d3}
J.~Zhang, C.-X. Zhang, Y.~Liu, Y.-X. Jin, X.-W. Yang, B.~Zheng, Y.~Liu, and L.-Z. Guo, ``D3: Diversity, difficulty, and dependability-aware data selection for sample-efficient llm instruction tuning,'' \emph{arXiv preprint arXiv:2503.11441}, 2025.

\bibitem[Gligori{\'c} et~al.(2024)Gligori{\'c}, Zrnic, Lee, Cand{\`e}s, and Jurafsky]{gligoric2024can}
K.~Gligori{\'c}, T.~Zrnic, C.~Lee, E.~J. Cand{\`e}s, and D.~Jurafsky, ``Can unconfident llm annotations be used for confident conclusions?'' \emph{arXiv preprint arXiv:2408.15204}, 2024.

\bibitem[{OpenAI}(2024)]{embedding_model}
\BIBentryALTinterwordspacing
{OpenAI}. (2024, jan) text-embedding-3-large. [Online]. Available: \url{https://platform.openai.com/docs/models/text-embedding-3-large}
\BIBentrySTDinterwordspacing

\bibitem[{liteLLM}(2025)]{litellm_docs}
\BIBentryALTinterwordspacing
{liteLLM}. (2025, Jun) Litellm documentation. [Online]. Available: \url{https://docs.litellm.ai/}
\BIBentrySTDinterwordspacing

\bibitem[{Reddit users}(2025)]{deepseekTimeout2025}
{Reddit users}, ``Reddit discusses deepseek api issues,'' \url{https://www.reddit.com/r/LocalLLaMA/comments/1ichohj/deepseek_api_every_request_is_a_timeout}, 2025, accessed: May 17, 2025.

\bibitem[{GitHub users}(2025)]{deepseekGitHub2025}
{GitHub users}, ``Github issue on deepseek api performance,'' \url{https://github.com/cline/cline/issues/1238}, 2025, accessed: May 17, 2025.

\bibitem[{BytePlus}(2025)]{byteplusInfra2025}
{BytePlus}, ``Byteplus discussion on infrastructure constraints,'' \url{https://www.byteplus.com/en/topic/375032}, 2025, accessed: May 17, 2025.

\bibitem[{Milvus}(2025)]{milvusLatency2025}
{Milvus}, ``Milvus latency metrics for deepseek r1,'' \href{https://milvus.io/ai-quick-reference/what-is-the-latency-of-deepseeks-r1-model-in-production-environments}{Milvus Latency Metrics for DeepSeek R1}, 2025, accessed: May 17, 2025.

\bibitem[{OpenAI}(2025)]{openaipricing}
\BIBentryALTinterwordspacing
{OpenAI}, ``{Pricing - OpenAI API},'' 2025, accessed: 2025-05-15. [Online]. Available: \url{https://platform.openai.com/docs/pricing}
\BIBentrySTDinterwordspacing

\bibitem[{DeepSeek}(2025)]{deepseekpricing}
\BIBentryALTinterwordspacing
{DeepSeek}, ``Models \& pricing | deepseek api docs,'' 2025, accessed: 2025-05-15. [Online]. Available: \url{https://api-docs.deepseek.com/quick_start/pricing}
\BIBentrySTDinterwordspacing

\bibitem[Brown et~al.(2020)Brown, Mann, Ryder, Subbiah, Kaplan, Dhariwal, Neelakantan, Shyam, Sastry, Askell, et~al.]{brown2020language}
T.~Brown, B.~Mann, N.~Ryder, M.~Subbiah, J.~D. Kaplan, P.~Dhariwal, A.~Neelakantan, P.~Shyam, G.~Sastry, A.~Askell \emph{et~al.}, ``Language models are few-shot learners,'' \emph{Advances in neural information processing systems}, vol.~33, pp. 1877--1901, 2020.

\bibitem[Rathje et~al.(2024)Rathje, Mirea, Sucholutsky, Marjieh, Robertson, and Van~Bavel]{rathje2024gpt}
S.~Rathje, D.-M. Mirea, I.~Sucholutsky, R.~Marjieh, C.~E. Robertson, and J.~J. Van~Bavel, ``Gpt is an effective tool for multilingual psychological text analysis,'' \emph{Proceedings of the National Academy of Sciences}, vol. 121, no.~34, p. e2308950121, 2024.

\bibitem[Kolla et~al.(2024)Kolla, Salunkhe, Chandrasekharan, and Saha]{kolla2024llm}
M.~Kolla, S.~Salunkhe, E.~Chandrasekharan, and K.~Saha, ``Llm-mod: Can large language models assist content moderation?'' in \emph{Extended Abstracts of the CHI Conference on Human Factors in Computing Systems}, 2024, pp. 1--8.

\bibitem[Savelka and Ashley(2023)]{savelka2023unreasonable}
J.~Savelka and K.~D. Ashley, ``The unreasonable effectiveness of large language models in zero-shot semantic annotation of legal texts,'' \emph{Frontiers in Artificial Intelligence}, vol.~6, p. 1279794, 2023.

\bibitem[Sushil et~al.(2024)Sushil, Zack, Mandair, Zheng, Wali, Yu, Quan, Lituiev, and Butte]{sushil2024comparative}
M.~Sushil, T.~Zack, D.~Mandair, Z.~Zheng, A.~Wali, Y.-N. Yu, Y.~Quan, D.~Lituiev, and A.~J. Butte, ``A comparative study of large language model-based zero-shot inference and task-specific supervised classification of breast cancer pathology reports,'' \emph{Journal of the American Medical Informatics Association}, vol.~31, no.~10, pp. 2315--2327, 2024.

\bibitem[Yu et~al.(2019)Yu, Theisen, Williams, and Menzies]{yu2019improving}
Z.~Yu, C.~Theisen, L.~Williams, and T.~Menzies, ``Improving vulnerability inspection efficiency using active learning,'' \emph{IEEE Transactions on Software Engineering}, vol.~47, no.~11, pp. 2401--2420, 2019.

\bibitem[Sazid et~al.(2024)Sazid, Kuri, Ahmed, and Satter]{sazid2024commit}
Y.~Sazid, S.~Kuri, K.~S. Ahmed, and A.~Satter, ``Commit classification into maintenance activities using in-context learning capabilities of large language models.'' in \emph{ENASE}, 2024, pp. 506--512.

\bibitem[Sheikhaei et~al.(2024)Sheikhaei, Tian, Wang, and Xu]{sheikhaei2024empirical}
M.~S. Sheikhaei, Y.~Tian, S.~Wang, and B.~Xu, ``An empirical study on the effectiveness of large language models for satd identification and classification,'' \emph{Empirical Software Engineering}, vol.~29, no.~6, p. 159, 2024.

\bibitem[Lambert et~al.(2024)Lambert, Ishitani, and Xavier]{lambert2024identification}
P.~Lambert, L.~Ishitani, and L.~Xavier, ``On the identification of self-admitted technical debt with large language models,'' in \emph{Simp{\'o}sio Brasileiro de Engenharia de Software (SBES)}.\hskip 1em plus 0.5em minus 0.4em\relax SBC, 2024, pp. 651--657.

\bibitem[Arshad et~al.(2024)Arshad, Riaz, Fatima, and Yasin]{arshad2024sevpredict}
M.~A. Arshad, A.~Riaz, R.~Fatima, and A.~Yasin, ``Sevpredict: Exploring the potential of large language models in software maintenance,'' \emph{AI}, vol.~5, no.~4, pp. 2739--2760, 2024.

\bibitem[Zhang et~al.(2023)Zhang, Irsan, Thung, and Lo]{zhang2023cupid}
T.~Zhang, I.~C. Irsan, F.~Thung, and D.~Lo, ``Cupid: Leveraging chatgpt for more accurate duplicate bug report detection,'' \emph{arXiv preprint arXiv:2308.10022}, 2023.

\bibitem[Gerosa et~al.(2024)Gerosa, Trinkenreich, Steinmacher, and Sarma]{Gerosa2024}
\BIBentryALTinterwordspacing
M.~Gerosa, B.~Trinkenreich, I.~Steinmacher, and A.~Sarma, ``Can ai serve as a substitute for human subjects in software engineering research?'' \emph{Automated Software Engineering}, vol.~31, no.~1, p.~13, 2024. [Online]. Available: \url{https://doi.org/10.1007/s10515-023-00409-6}
\BIBentrySTDinterwordspacing

\bibitem[Zhang et~al.(2025{\natexlab{c}})Zhang, He, Zhang, Kang, Li, Xie, Wang, Wang, Huang, Fu, Nallipogu, Lin, Dang, Rajmohan, and Zhang]{zhang2025swebenchgoeslive}
\BIBentryALTinterwordspacing
L.~Zhang, S.~He, C.~Zhang, Y.~Kang, B.~Li, C.~Xie, J.~Wang, M.~Wang, Y.~Huang, S.~Fu, E.~Nallipogu, Q.~Lin, Y.~Dang, S.~Rajmohan, and D.~Zhang, ``Swe-bench goes live!'' 2025. [Online]. Available: \url{https://arxiv.org/abs/2505.23419}
\BIBentrySTDinterwordspacing

\bibitem[Wang et~al.(2025)Wang, Pradel, and Liu]{wang2025solved}
Y.~Wang, M.~Pradel, and Z.~Liu, ``Are" solved issues" in swe-bench really solved correctly? an empirical study,'' \emph{arXiv preprint arXiv:2503.15223}, 2025.

\bibitem[Miserendino et~al.(2025)Miserendino, Wang, Patwardhan, and Heidecke]{miserendino2025swe}
S.~Miserendino, M.~Wang, T.~Patwardhan, and J.~Heidecke, ``Swe-lancer: Can frontier llms earn \$1 million from real-world freelance software engineering?'' \emph{arXiv preprint arXiv:2502.12115}, 2025.

\bibitem[Wang et~al.(2024)Wang, Li, Song, Xu, Tang, Zhuge, Pan, Song, Li, Singh, Tran, Li, Ma, Zheng, Qian, Shao, Muennighoff, Zhang, Hui, Lin, Brennan, Peng, Ji, and Neubig]{openhands}
\BIBentryALTinterwordspacing
X.~Wang, B.~Li, Y.~Song, F.~F. Xu, X.~Tang, M.~Zhuge, J.~Pan, Y.~Song, B.~Li, J.~Singh, H.~H. Tran, F.~Li, R.~Ma, M.~Zheng, B.~Qian, Y.~Shao, N.~Muennighoff, Y.~Zhang, B.~Hui, J.~Lin, R.~Brennan, H.~Peng, H.~Ji, and G.~Neubig, ``{OpenHands: An Open Platform for AI Software Developers as Generalist Agents},'' 2024. [Online]. Available: \url{https://arxiv.org/abs/2407.16741}
\BIBentrySTDinterwordspacing

\end{thebibliography}
\end{small}

\end{document}